\documentclass[aps,twocolumn,floats,showpacs]{revtex4}
\usepackage{graphicx}

\begin{document}
\title{Numerical study of the disorder-driven roughening transition in an\\
elastic manifold in a periodic potential}

\author{Jae Dong Noh}
\affiliation{Theoretische Physik, Universit\"{a}t des Saarlandes,
66041 Saarbr\"{u}cken, Germany}
\author{Heiko Rieger}
\affiliation{Theoretische Physik, Universit\"{a}t des Saarlandes,
66041 Saarbr\"{u}cken, Germany}

\date{\today}

\begin{abstract}
  We investigate the roughening phase transition of a
  $(3+1)$-dimensional elastic manifold driven by the completion between
  a periodic pinning potential and a randomly distributed impurities.
  The elastic manifold is modeled by a solid-on-solid type interface
  model, and universal features of the transition from a flat phase
  (for strong periodic potential) to a rough phase (for strong
  disorder) are studied at zero temperature using a combinatorial
  optimization algorithm technique. We find a {\it continuous}
  transition with a set of numerically estimated critical exponents
  that we compare with analytic results and those for a periodic
  elastic medium.
\end{abstract}

\maketitle

Extended objects embedded in a higher-dimensional space, like
polymers, magnetic flux lines, surfaces, interfaces or domain walls,
are on large length scales commonly described by elastic models, the
so-called elastic manifolds (EM) \cite{Halpin-Healy95}. If quenched
disorder in form of impurities or other randomly distributed pinning
centers is present, the EM will be, even at low temperatures or in the
absence of any thermal fluctuations, in a rough state. However, a
periodic array of pinning potentials, like a background lattice
potential, increases the tendency of the EM to minimize their elastic
energy, i.e.~to stay flat \cite{bouchaud92,nattermann}. Both
mechanism compete with each other and by varying their relative
strength a roughening transition might emerge. The numerical
investigation of such a scenario is the purpose of this paper.

Consider a $d$-dimensional EM in a $(d+1)$-dimensional medium with
quenched impurities distributed randomly.  Fluctuations of the shape
of the EM are then described by a {\em scalar} displacement field
$\phi({\bf r})$ denoting a deviation from a flat reference state in
the $(d+1)$th direction at each ${\bf r}\in {\bf R}^d$; $(\phi({\bf
r}),{\bf r})$ refers to the $(d+1)$-dimensional coordinate of an EM
segment.  It is known that quenched disorder, no matter how weak,
destabilizes the flat phase for $d<4$~\cite{larkin70}.  The emerging
disordered rough phase is characterized by a divergent displacement
correlation function $B({\bf r}) \equiv \overline{ \langle [\phi({\bf
r}) - \phi({\bf 0})]^2 \rangle}$ with the distance $r=|{\bf r}|$, whose
scaling property is universal.

An example of the 1D EM is a directed polymer or a magnetic flux line in 
a disordered 2D plane. Using a mapping to the Kardar-Parisi-Zhang equation 
for surface growth, it is shown analytically that the displacement
correlation diverges algebraically as 
\begin{equation}\label{power-law}
B({\bf r}) \sim r^{2\zeta}
\end{equation} 
with the exactly-known roughness exponent 
$\zeta=\frac{2}{3}$~\cite{Halpin-Healy95}.
For higher-dimensional EMs, analytic studies using 
a functional renormalization group~(FRG) method predict that the 
rough phase is governed by a zero-temperature fixed point which is also 
characterized by a power-law divergence of $B({\bf r})$
as in Eq.~(\ref{power-law})~\cite{fisher86,halpin-healy90} and 
the roughness exponent is found to be $\zeta = 0.2083\epsilon
+{\cal O}(\epsilon^2)$ \cite{fisher86} or 
$\frac{2}{9}\epsilon + {\cal O}(\epsilon^2)$ 
\cite{halpin-healy90} up to first order in $\epsilon=4-d$.
These values are in good agreement with numerical estimates obtained
from exact ground-state calculations for $(2+1)$ and $(3+1)$ D
solid-on-solid type interface models~\cite{alava96,middleton95}.

In addition to random impurities, also the structure of the embedding
medium can affect the large scale properties of the EM. In particular,
if the medium has a {\em crystalline} structure, the EM is pinned by
the disorder potential and by the periodic crystal potential, both
effects competing with each other --- the latter favoring a flat state
while the former a rough one. Hence, the EM may undergo a phase
transition at a critical disorder-potential strength.  Following a
qualitative perturbative scaling argument~\cite{nattermann}, one can
see that there exists the disorder-driven roughening transition at
nonzero disorder strength for $2<d<4$.

In 2D, the disorder potential is argued
to dominate over the periodic potential
marginally~\cite{bouchaud92,nattermann}. Consequently 
the 2D EM is believed to be rough at any disorder strength asymptotically
beyond a certain length scale which diverges exponentially with the 
inverse of a disorder strength. 
Some numerical studies support such claim~\cite{seppala01}. On the
other hand, it has been reported that an Ising domain wall in 
a (2+1)D lattice with bond dilution displays a roughening transition 
at a non-zero dilution probability~\cite{alava96}. 
It suggests that the type of disorder might be important in the marginal 
2D case~(see also discussions in Ref.~\cite{seppala01}). 

In 3D, the existence of a roughening transition was shown in the
studies using a Gaussian variational~(GV) method~\cite{bouchaud92} and
a FRG method~\cite{emig98,nattermann}.  In the GV study, the free
energy was calculated by approximating the Hamiltonian of the EM with
a Gaussian. It leads to a conclusion that the transition is of first
order.  On the other hand, the FRG study with a perturbative expansion
in the periodic potential strength and in $\epsilon=4-d$ yields that
the transition is continuous with a correlation length exponent $\nu =
1/(2\sqrt{\epsilon})$.  To clarify this issue we performed in this
work a numerical study of the disorder-driven roughening transition of
the 3D elastic manifold in a crystal potential with quenched random
impurities.

The EM is described by the Hamiltonian~\cite{nattermann}
\begin{equation}\label{Hamiltonian}
{\cal H} = \int d^d {\bf r} \left[
\frac{\gamma}{2} | \nabla \phi({\bf r}) |^2  + V_P (\phi({\bf r}))
+ V_R(\phi({\bf r}),{\bf r}) \right] \ ,
\end{equation}
where the first term represents a surface tension, 
$V_P(\phi) = -V \cos \phi$ the periodic lattice potential,
and $V_R(\phi,{\bf r})$ the quenched random potential. Here $\phi$ is
measured in units of $a_0/(2\pi)$ with $a_0$ the lattice constant of the
crystal. For uncorrelated distribution of impurities, 
$V_R(\phi,{\bf r})$ can be taken as a random variable with mean zero 
and variance given by
\begin{equation}\label{correlator}
\overline{ V_R(\phi,{\bf r}) V_R(\phi',{\bf r'}) } =
D^2 R(\phi-\phi')\ \delta({\bf r}-{\bf r'})
\end{equation}
with a parameter $D$ for the disorder strength.
Uncorrelated distribution of impurities implies that the disorder
correlation function in the $(d+1)$th direction is also {\em short-ranged}, 
i.e., $R(\phi) = \delta(\phi)$.

It is interesting to note that the Hamiltonian in Eq.~(\ref{Hamiltonian})
can also describe so-called {\em periodic 
elastic media}~(PEM) in a crystal with quenched disorder~\cite{nattermann}. 
A system of strongly interacting  
particles or other objects, like magnetic flux lines in a type-II 
superconductor or a charge density in a solid, will order at low 
temperatures into a regular arrangement, namely, the flux line 
lattice~(FLL) or the charge density wave~(CDW), respectively.
Fluctuations either induced by thermal noise or by disorder induce 
deviations of the
individual particles from their equilibrium positions. As long as
these fluctuations are not too strong, an expansion of the interaction
energy around these equilibrium configuration might be appropriate. 
An expansion up to second order leads to the surface-tension-like 
term as in Eq.~(\ref{Hamiltonian}). In contrast to the EM,
the PEM have their own periodicity $\lambda$. 
It implies that the disorder potential 
$V_R(\phi,{\bf r})$ should be a periodic function in $\phi$ with
the periodicity $p=\frac{\lambda}{a_0}$ (commensurability parameter), 
even though the impurities are distributed randomly. Hence, the disorder
correlation function in Eq.~(\ref{correlator}) should be periodic: 
$R(\phi+2\pi p) = R(\phi)$. 
In 3D, as a result of the periodicity, 
the displacement correlation function for $\varphi\equiv\phi/p$ 
diverges logarithmically as
$ \overline{ \langle[ \varphi({\bf r}) - \varphi({\bf 0})]^2 \rangle}
\simeq 2 A \ln r$
with a universal coefficient $A\simeq 1.0$ in
the rough phase~\cite{giamarchi,mcnamara99,noh01}.
The periodic elastic media also display a disorder-driven 
roughening transition as a result of competition between the 
periodic potential and the random potential. 
However, there is a slight controversy regarding the nature
of the transition since an analytic FRG study~\cite{nattermann97} and
a zero-temperature numerical study~\cite{noh01} yield results that 
are not fully compatible.

We introduce a discrete solid-on-solid~(SOS) type interface model for 
the elastic manifold whose continuum Hamiltonian is given in 
Eq.~(\ref{Hamiltonian}). Locally the EM remains flat in one of 
periodic potential minima at $\phi = 2\pi h$ with integer $h$.  
Due to fluctuations, some regions might shift to a different
minimum with another value of $h$ to create a step (or domain
wall) separating domains. To minimize the cost of the elastic 
and periodic potential energy in Eq.~(\ref{Hamiltonian}), 
the domain-wall width must be finite, say $\xi_o$~\cite{nattermann}.
Therefore, if one neglects fluctuations in length scales less than $\xi_o$, the
continuous displacement field $\phi({\bf r})$ can be replaced by the
integer height variable $\{h_{\bf x}\}$ representing a $(3+1)$ D
SOS interface on a simple cubic lattice with sites ${\bf x}\in
\{1,\ldots,L\}^3$. 
The lattice constant is of order $\xi_o$ and set to
unity. The energy of the interface is given by 
the Hamiltonian
\begin{equation}\label{H_SOS}
{\cal H} = \sum_{\langle{\bf x,y}\rangle} 
J_{(h_{\bf x},{\bf x}); (h_{\bf y},{\bf y})}
| h_{\bf x} - h_{\bf y} |
- \sum_{{\bf x}} V_R (h_{\bf x},{\bf x}) \ ,
\label{ham-dis}
\end{equation}
where the first sum is over nearest neighbor site pairs.
After the coarse graining, the step energy $J>0$ as well as the
random pinning potential energy $V_R$ becomes a quenched random
variable distributed independently and randomly. 
Note the PEM has the same Hamiltonian as in Eq.~(\ref{H_SOS}) 
with random but periodic $J$ and $V_R$ in $h$ with periodicity 
$p$~\cite{noh01}. In this sense, the elastic manifold emerges as
in the limit $p\rightarrow \infty$ of the periodic elastic medium.

Here, we are interested only in the ground state property. Since the
prevalent RG-picture suggests that the roughening transition is
described by a zero temperature fixed point
\cite{emig98,giamarchi,nattermann} one expects the critical exponents
we find to be valid for the finite temperature roughening transition
as well. To find the ground state, one maps the 3D SOS model onto a
ferromagnetic random bond Ising model in $(3+1)$D hypercubic lattice
with anti-periodic boundary conditions in the extra dimension
\cite{middleton95} (for the 3 space direction one uses periodic 
boundary conditions instead). The anti-periodic boundary conditions force a
domain wall into the ground state configuration of the (3+1)D
ferromagnet.  Note that bubbles are {\it not} present in the ground
state.  A domain wall may contain an overhang which is unphysical in
the interface interpretation. Fortunately, one can forbid overhangs in
the Ising model representation using a technique described in
Ref.~\cite{middleton95}.  If the longitudinal and transversal bond
strengths are assigned with $J/2$ and $V_R/2$ occurring in
Eq.~(\ref{ham-dis}), respectively, this domain wall of the
ferromagnet becomes equivalent to the ground state configuration of
(\ref{ham-dis}) for the interface with the same energy. The domain
wall with the lowest energy is then determined exactly by using a
combinatorial optimization algorithm, a so-called max-flow/min-cost
algorithm.  This combinatorial optimization technique is nowadays
standard in the study of disordered systems and we refer readers to
Ref.~\cite{heiko} for a detailed review.

We performed the ground state calculation on $L^3\times H$
hypercubic lattices for $L\leq 32$. $H$, the size in the extra
direction, is taken to be larger than the interface width.
Several distributions for $J$ and $V_R$ were studied for the critical
behavior of a disordered system may depend on the choice of the
disorder distribution as in the random field Ising system~\cite{hartmann99}. 
However,
our main numerical results do not depend on the specific choice of the
distribution. So we only present the results for an exponential
distribution for $J>0$, $P(J) = e^{-J/J_0} / J_0$ and uniform
distribution for $0\leq V_R \leq V_{max}$. 
The disorder strength is controlled with the parameter 
$\Delta \equiv V_{max}/{J_0}$.
Other distributions
studied include (bimodal,bimodal) and (uniform,uniform) distributions
for $(J,V_R)$, and gave identical estimates for the critical exponents.

The state of the interface is characterized by the width $W$
defined as $W^2 = \overline{ \langle h_{\bf x}^2\rangle_o -
\langle h_{\bf x}\rangle_o^2 }$ , where $\langle\cdots\rangle_o$
denotes the spatial average in the ground state and
$\overline{\cdots}$ the disorder average. $W^2$ is proportional to the
spatial integral of the displacement correlation function. We also
measure a magnetization-like quantity $m \equiv \overline{ | \langle
e^{i\pi h_{\bf x}}\rangle_o | }$. It is analogous to the magnetization
used as an order parameter for the roughening transition of the
PEM~\cite{noh01}.  One expects that $m$ is non-zero in the flat phase
and vanishes in the rough phase.  So it can be used as an order
parameter for the roughening transition.

\begin{figure}
\includegraphics[width=\columnwidth]{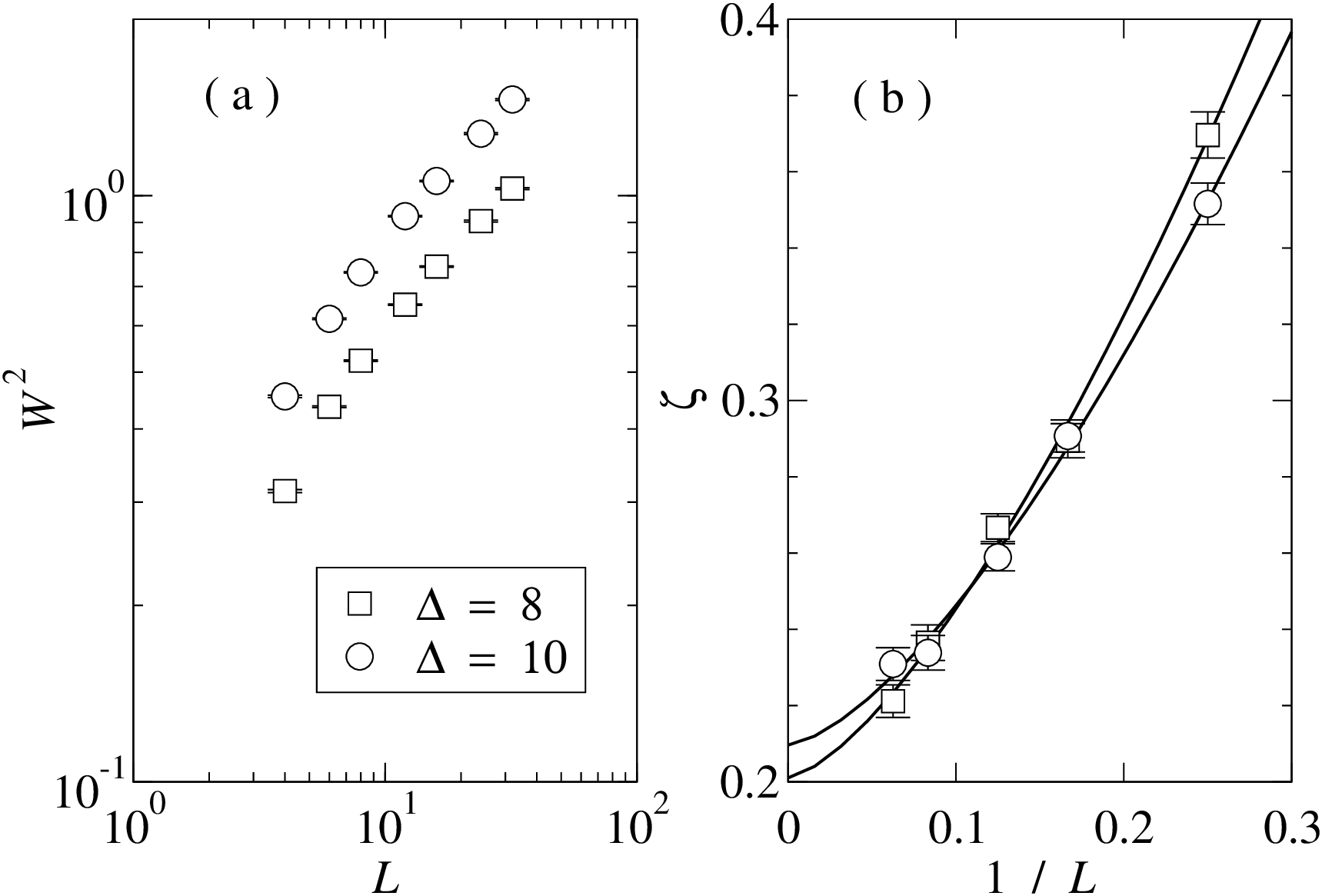}
\caption{\label{W_rough} (a) $W$ vs. $L$ in a log-log plot
  at $\Delta=8$ and $10$ (strong disorder).  (b) $\zeta$ vs. $1/L$
  for the same values of $\Delta$ as in (a).
  The solid lines are least square fits to the form $\zeta(L) = \zeta
  + a L^{-b}$.  We obtain $\zeta=0.21\pm0.01$ and the resulting value
  of $b\simeq 1.5 >1$ indicates that the extrapolation is stable
  against statistical uncertainties.}
\end{figure}
We first examine the power-law scaling behavior of the width, $W\sim
L^\zeta$, in the rough phase 
at large disorder strength $\Delta=8$ and $10$.
Figure~\ref{W_rough} (a) shows the width average over $1000\sim 5000$
disorder realizations for $L=4\sim 32$. A noticeable curvature 
in the log-log plot indicates that corrections to scaling
are still rather strong. Nevertheless, we can estimate the
roughness exponent by extrapolating an effective exponent
$\zeta(L)=\log[W(2L)/W(L)]/\log 2$ to the limit $L\rightarrow \infty$ by
fitting it to a form $\zeta(L) = \zeta + a L^{-b}$, see 
Fig.~\ref{W_rough} (b).
We obtain that
\begin{equation}
\zeta = 0.21 \pm 0.01 \ .
\end{equation}
This value is consistent with the previous analytical and numerical
results~\cite{fisher86,halpin-healy90,middleton95,alava96}.
\begin{figure}
\includegraphics[width=\columnwidth]{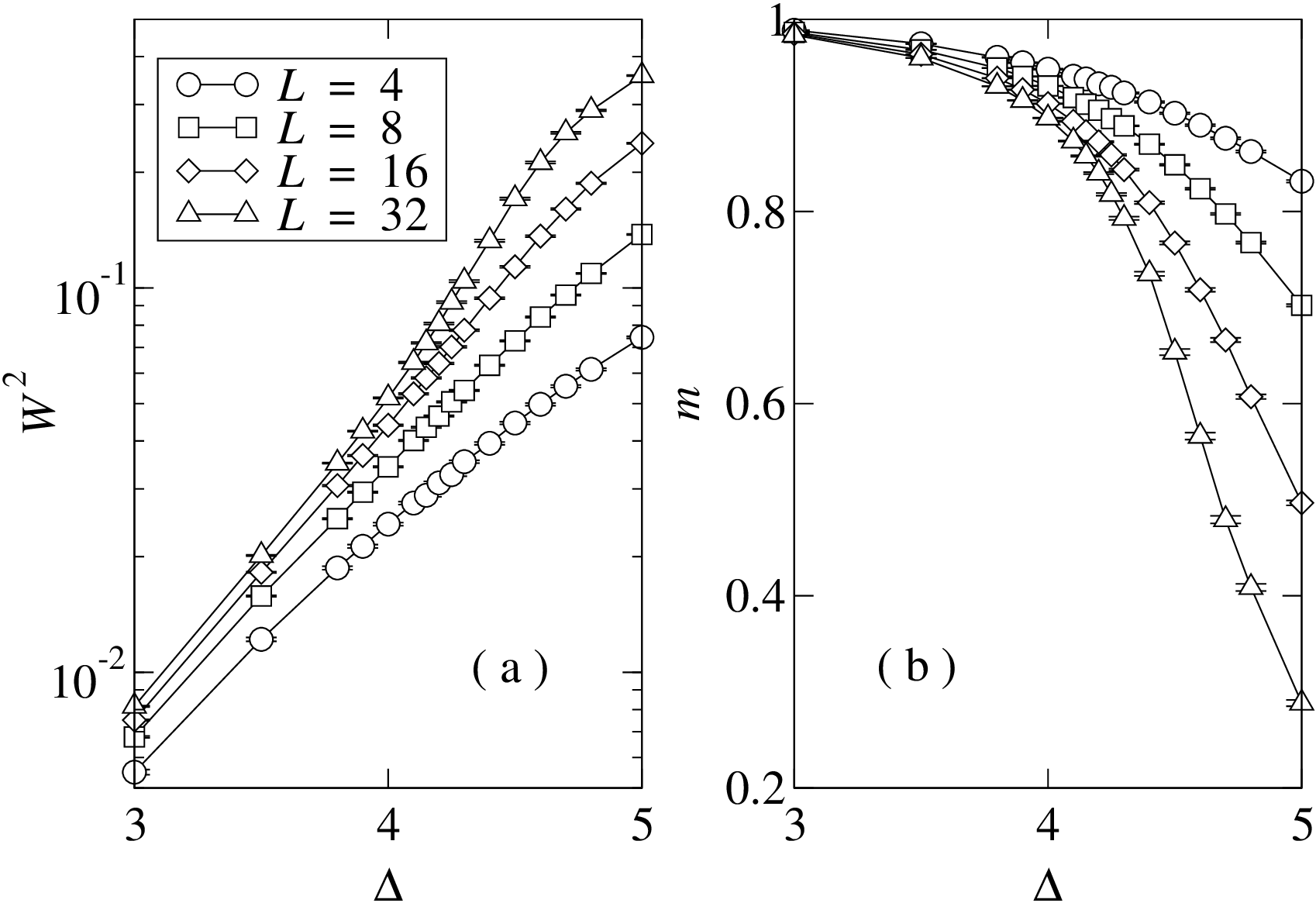}
\caption{\label{W_M}The interface width (a) and the order parameter
(b) as functions of $\Delta$ at $L=4,8,16$, and $32$. They are
obtained from the disorder average over $3000\sim 20000$ samples.}
\end{figure}

\begin{figure}
\includegraphics[width=\columnwidth]{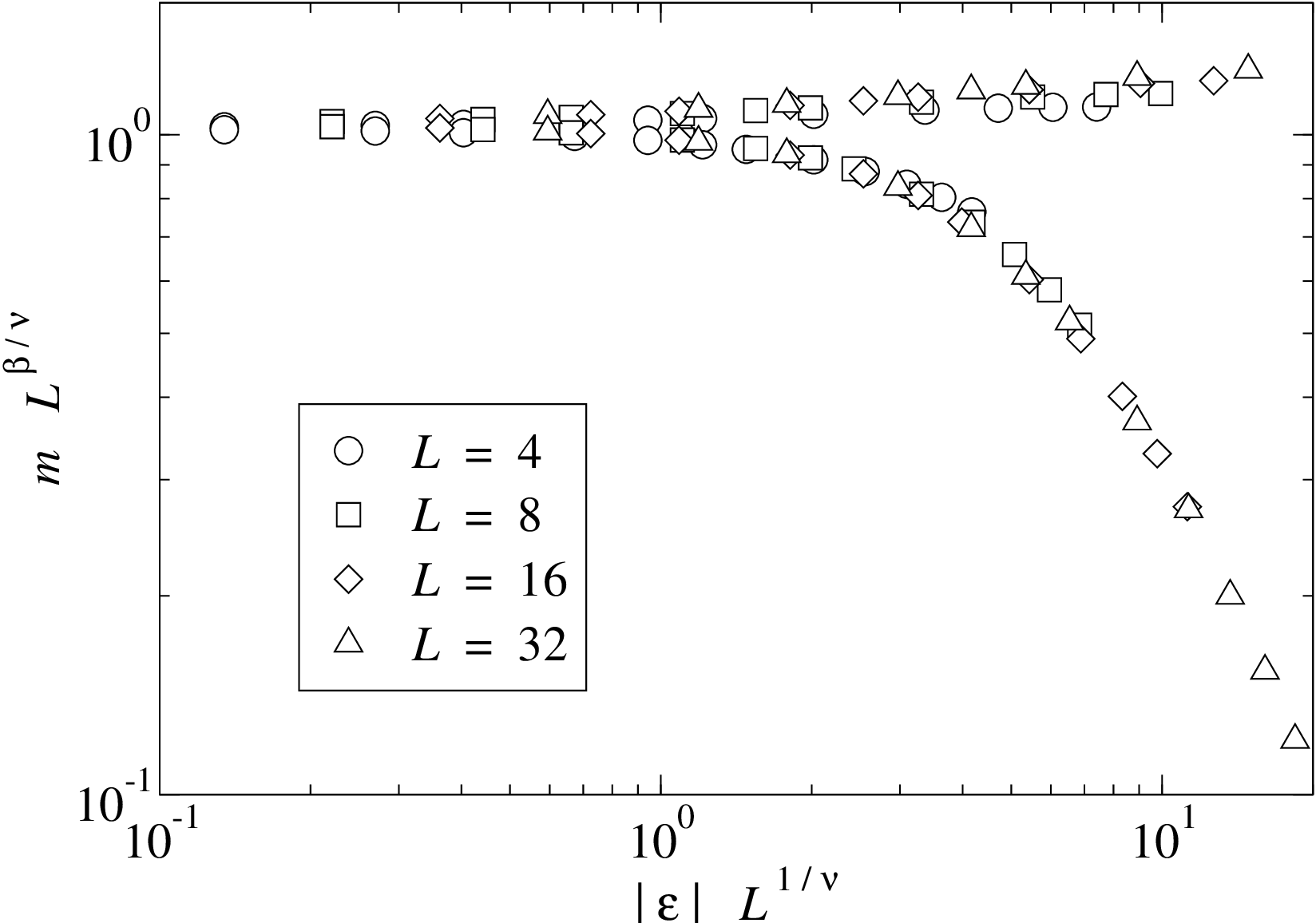}
\caption{\label{m_scale} Scaling plot of $m L^{\beta/\nu}$ vs.
$|\varepsilon| L^{1/\nu}$ with $\varepsilon=\Delta-4.25$, 
$\beta/\nu=0.07$, and $\nu=1.4$.}
\end{figure}

As the disorder strength decreases, the width also decreases and the
interface eventually becomes flat below a threshold~(see
Fig.~\ref{W_M}). The order parameter also shows an indication of
a phase transition around $\Delta\simeq 4.0$. 
Apparently the order parameter decreases {\em continuously}.
Therefore we perform a scaling analysis assuming that the phase transition
is a continuous one. 
The critical point $\Delta_c$ can be determined from the
finite-size-scaling property of the order parameter:
\begin{equation}\label{m_scaling}
m(L,\varepsilon) = L^{-\beta/\nu} {\cal F}( \varepsilon L^{1/\nu})  \ ,
\end{equation}
where $\varepsilon\equiv \Delta-\Delta_c$, and $\beta$~($\nu$) is the
order parameter~(correlation length) exponent.
The scaling function ${\cal F}(x)$ has a limiting behavior ${\cal
F}(x\rightarrow 0) = \mbox{const.}$ so that the order parameter decays
algebraically with $L$ as $m\sim L^{-\beta/\nu}$ at the critical point.
It also behaves as ${\cal F}(x\rightarrow -\infty)
\sim |x|^{\beta}$ so that $m\sim |\varepsilon|^{\beta}$ for $\Delta <\Delta_c$
in the infinite system size limit.
\begin{figure}
\includegraphics[width=\columnwidth]{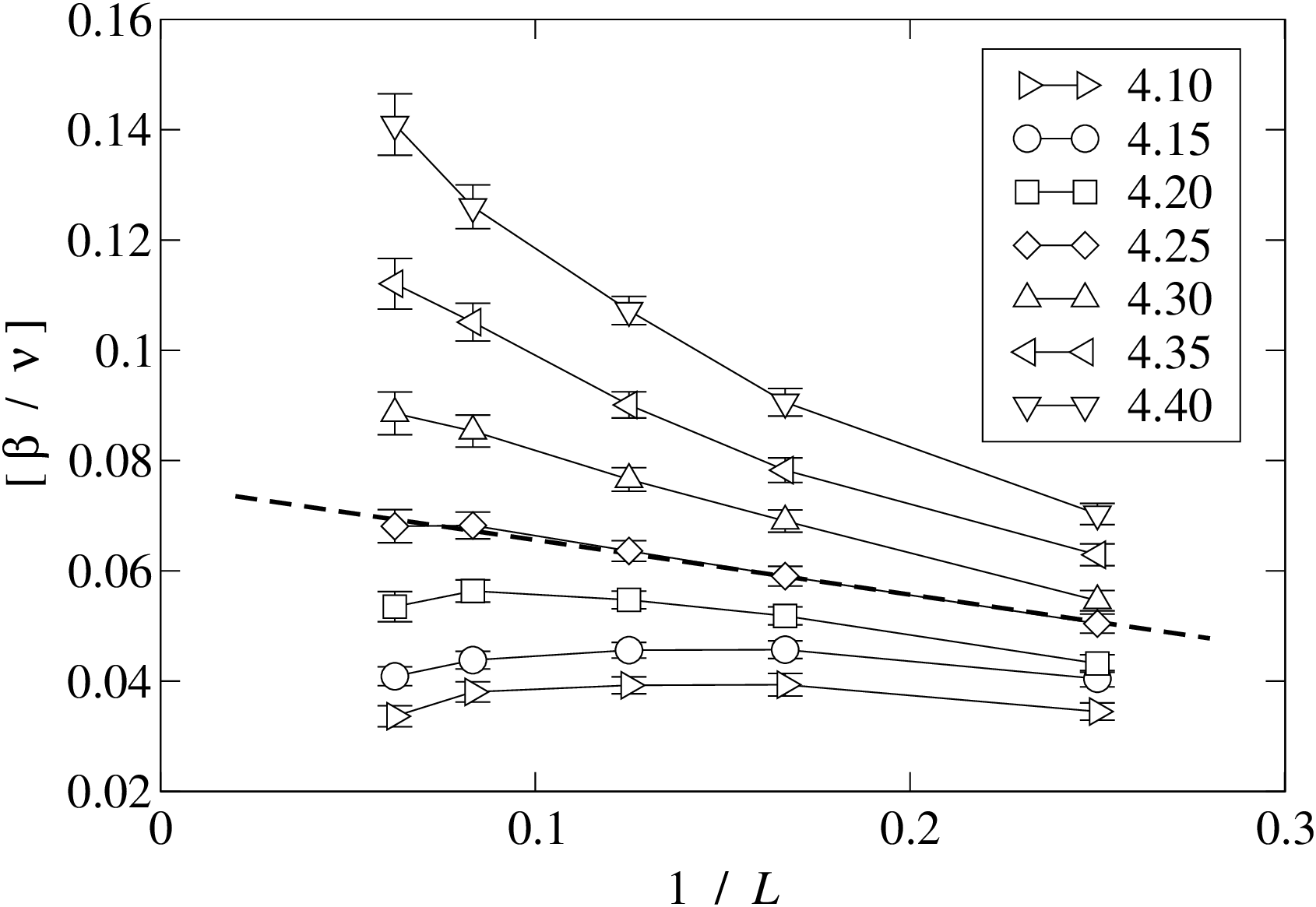}
\caption{\label{beta_nu} The effective exponent 
  $[\beta/\nu](L)$ for different values of $\Delta$ as a function of
  $1/L$. The broken line is a guide for the eyes that separates the
  curves with a downward bending ($\Delta<\Delta_c$) from those with
  an upward bending ($\Delta>\Delta_c$).}
\end{figure}
Consider the effective exponent 
$[\beta/\nu](L) = - \log( m(2L)/m(L)) / \log 2$. 
It converges to the value of $\beta/\nu$ at the critical point 
and deviates from it otherwise as $L$ increases.
We estimate the critical threshold as the optimal value of $\Delta$ at which
the effective exponent approaches a nontrivial value. The plot for
this effective exponent is shown in Fig.~\ref{beta_nu}. One can see
that there is a downward and upward curvature for $\Delta < 4.20$
and $\Delta > 4.30$, respectively. From this behavior we estimate
that $\Delta_c = 4.25 \pm 0.05$ and
\begin{equation}
\frac{\beta}{\nu} = 0.07 \pm 0.03 \ .
\end{equation}
Note that the effective exponent varies with $L$ even at the estimated
critical point, which implies that corrections to scaling are not 
negligible for system sizes up to $L=32$. 
For that reason our numerical results for
$\Delta_c$ and $\beta/\nu$ have rather large error bars, and one may
need larger system sizes for better precision. 
The exponents $\beta$ and $\nu$ could also be obtained from the scaling analysis
using Eq.~(\ref{m_scaling}). We fix the values of $\Delta_c$ and
$\beta/\nu$ to the values obtained before and vary $\nu$ to have
an optimal data collapse. We obtain 
\begin{equation}
\nu = 1.4 \pm 0.2
\end{equation}
and the corresponding scaling plot is shown in Fig.~\ref{m_scale}. 

\begin{figure}
\includegraphics[width=\columnwidth]{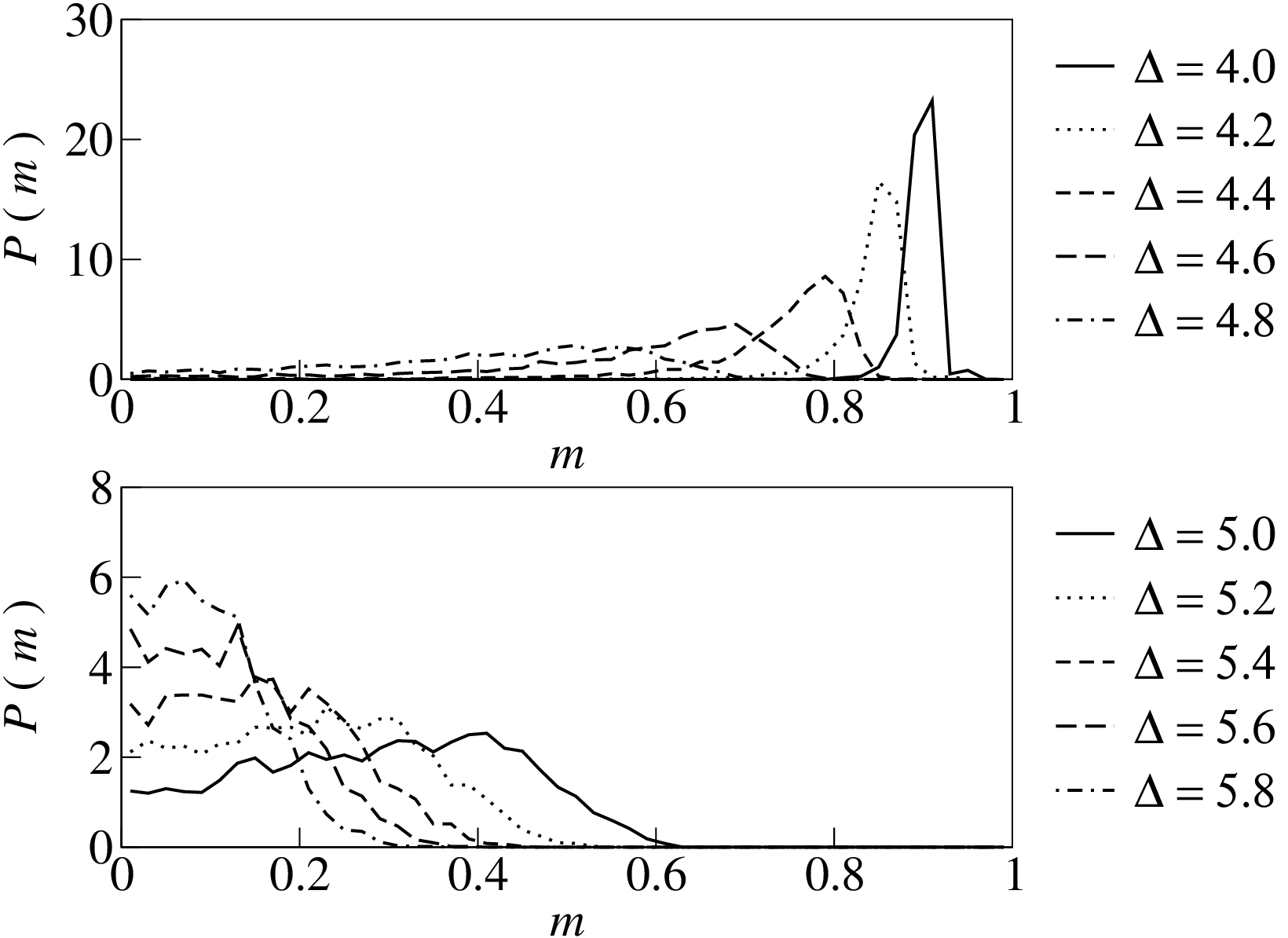}
\caption{\label{histogram} Histograms for the probability distribution
$P(m)$ of the magnetization at $L=32$.}
\end{figure}

The order-parameter scaling property shows that the roughening phase
transition is a {\em continuous} transition, though the exponent
$\beta \simeq 0.1$ is very small, as opposed to the results of the GV
study~\cite{bouchaud92} predicting a first order transition.  The
transition nature becomes more transparent by looking at the
probability distribution $P(m)$ of the magnetization near the critical
point. We measured the distribution from a histogram of the
magnetization of 3000 samples with $L=32$, which is shown in
Fig.~\ref{histogram}. We do not observe a double-peak structure in
$P(m)$, which would appear for a first order transition, at any values
of $\Delta$. Instead, there is a single peak which shifts continuously
toward zero as $\Delta$ increases.  We did not observe any double-peak
structure in the distribution of the width, either. Therefore we
conclude that the roughening transition is the continuous transition.

We note that this behavior is reminiscent of the three-dimensional
random field Ising-model~(RFIM)~\cite{rfim}, where also a very small
order is found. This causes an extremely weak systems size dependence
and the transition appears to be discontinuous in the order parameter
although the system is indeed critical and the correlation length
diverges. However, as has been pointed out in \cite{sourlas} the
magnetization of an {\it individual} sample shows close to the
transition discontinuous jumps when (for a fixed field distribution)
the coupling strength is varied (however, the sample-averaged
magnetization is smooth).  Since in the 3d RFIM the maximum jump does
not vanish even in the infinite size limit some objection against the
continuity of the phase transition has been raised \cite{sourlas}.
Since we find also a small order parameter exponent -- and because of
some model specific similarities between the 3d RFIM and the system we
study here --- we now want to check the jump size statistics, too.

For a given realization of $J$ and $V_R$ in Eq.~(\ref{H_SOS}), we
measure the order parameter and its jumps when changing $V_R$ with a
global factor.  The interface may undergo two types of intermittence:
The average height $\langle h\rangle$ may jump with a vanishing
overlap of the interface configurations before and after the jump
(meaning that the whole manifold is in a new position, uncorrelated
with the previous one). Or a large domain type excitation may appear
with only a small change in $\langle h\rangle$.  The intermittent
behavior of (1+1) and (2+1)D interfaces were studied in great detail
in Ref.~\cite{seppala01}. We want to study how the interface at a
given average height roughens, therefore we only take into account the
domain type excitations with the change in $\langle h\rangle$ less
than one half when measuring the jump of the order parameter. 

\begin{figure}
\includegraphics[width=\columnwidth]{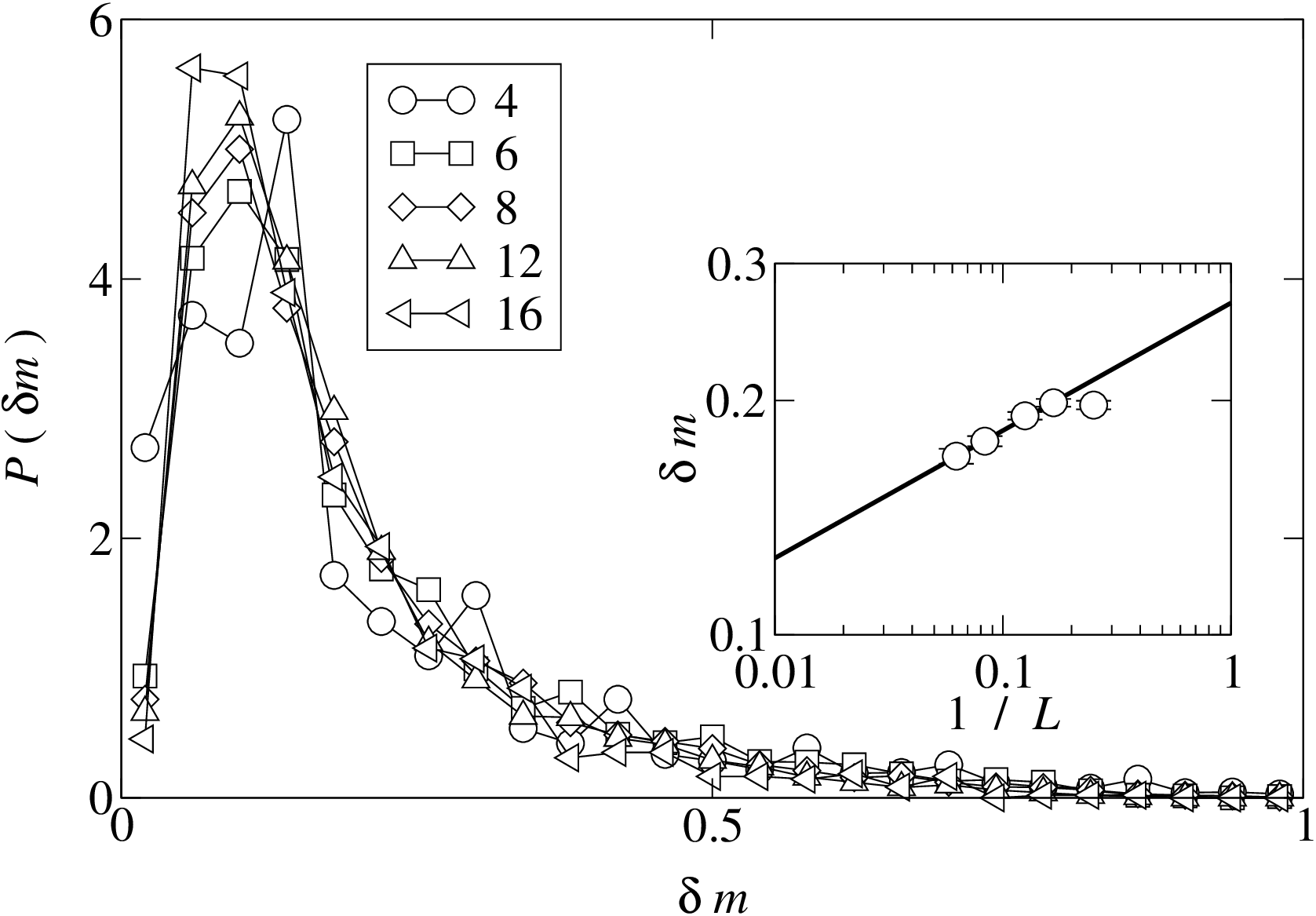}
\caption{\label{jump} Order parameter jump distribution. Inset
shows the plot of the averaged value vs. $1/L$ in the log-log scale. 
The solid line has a slope of 0.16.}
\end{figure}

We calculate the order parameter in the interval $3\leq \Delta \leq 7$
with spacing 0.02.  Figure~\ref{jump} shows the probability distribution
of {\em maximum} values of the jump $\delta m$ in $1000\sim 5000$ samples of
sizes $L=4,6,8,12,16$.  The inset shows the averaged value. 
It is non-zero, but becomes smaller as the
system size increases except for the case of $L=4$. The decay is very
slow, nevertheless we could fit the data for $L\ge 6$ to a form
$\delta m \sim L^{-0.16}$ as can be seen in the inset. It suggests
that the order parameter jump vanishes in the infinite size limit,
and the smallness of the estimated exponent $0.16$ is compatible
with our small estimate for the order parameter $\beta$.

\begin{figure}
\includegraphics[width=\columnwidth]{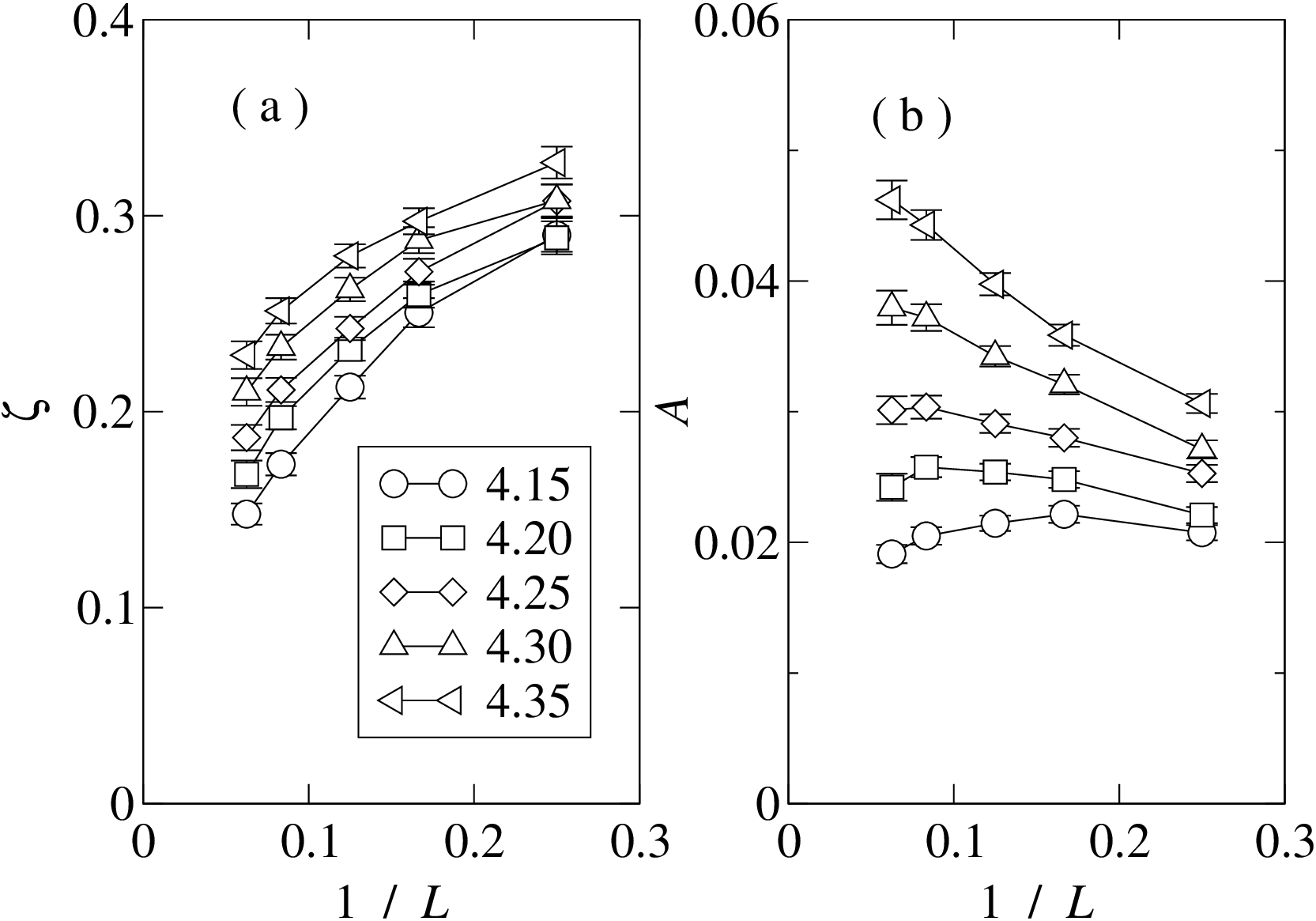}
\caption{\label{w_critical} (a) Effective exponent for the
power-law behavior of  $W\sim L^\zeta$. (b) Prefactor in
the logarithmic scaling of $W^2\simeq A\log L$.} 
\end{figure}

We also studied the scaling of the width at the critical
point. One might expect that the width at the critical point scales
as $W \sim L^{\zeta'}$ with a new roughness exponent $\zeta'$ different
from $\zeta\simeq 0.21$ for the rough interface. We plot the effective
exponent $\zeta=\log[W(2L)/W(L)] / \log2$ near the critical point in
Fig.~\ref{w_critical}~(a). One observes that this effective exponent does
not extrapolate to a non-zero value (for comparison c.f. 
Fig.~\ref{W_rough}~(b)).  
Instead, it decreases rapidly as $L$ grows, which suggests
rather a logarithmic scaling, $W^2 \simeq A \log L$, at the critical
point as in the case of the periodic elastic medium~\cite{noh01}. 
To investigate such a possibility, we measure the prefactor 
$A \equiv ( W^2(2L)-W^2(L)) / \log 2$ for this logarithmic scaling.  
They are also plotted in
Fig.~\ref{w_critical}~(b), which shows a clear threshold behavior: It
decreases~(increases) for $\Delta<4.25$~($>4.25$) and remains
constant~($A\simeq 0.03$) at $\Delta=4.25$, which was estimated as the
critical point from the order parameter scaling analysis.  These facts
are consistently suggesting that the interface width scales
logarithmically at the critical point:
\begin{equation}
W^2 \simeq 0.03 \log L \ .
\end{equation}
The result of the logarithmic scaling of the width at the critical point
agrees well with that of the FRG study~\cite{emig98}.

Our numerical results pose an interesting question: In recent work
\cite{noh01} we found that the disorder-driven roughening transition
of the PEM appears to be independent of the commensurability parameter
$p$~\cite{noh01}.  As mentioned before, the elastic manifold can be
seen as the $p\to\infty$ limit of the PEM. Based on this observation
one might speculate that the roughening transitions of both systems
belong to the same universality class. Although the $p\to\infty$-limit
of the PEM could also belong to a different universality class the
numerical results are compatible with the roughening transition in the
EM and the PEM belonging to the same universality class:: For the EM
we obtain $\beta/\nu \simeq 0.07$, $\nu \simeq 1.4$, and $A\simeq
0.03$~($W^2 \simeq A \log L$ at the critical point), while for the
PEM~\cite{noh01} we got $\beta/\nu \simeq 0.05$, $\nu \simeq 1.3$, and
$A\simeq 0.018$.  Although these values are very close to each other,
a final conclusion could not be drawn yet due to the rather strong
correction to finite size scaling observed in our study of the EM.  
So we have to leave this issue as an open question.

In summary, we have studied the $(3+1)$D elastic manifold in a crystal
with quenched random impurities. We have investigated numerically the
disorder-driven roughening transition at zero temperature using an
exact combinatorial optimization algorithm technique.  The transition
turns out to be continuous with the critical exponents $\beta/\nu
\simeq 0.07$ and $\nu \simeq 1.4$. For a given disorder potential
configuration, the order parameter shows a discrete jump for finite
size systems when varying its strength. However, the jump vanishes in
the infinite system size limit. We also found that $W^2$ scales
logarithmically with the system size $L$ at the critical point, in
contrast to the power-law scaling in the rough phase. Our results do
not agree with the scenario proposed in Ref.~\cite{bouchaud92} that
the roughening transition should be first order. Instead, they are in
a {\em qualitative} agreement with those of FRG $\epsilon$-expansion
study in Ref.~\cite{emig98}, which predicts a continuous
roughening transition and a logarithmic divergence of $W^2$ at the
critical point.  However, there is a significant discrepancy between
the values of the critical exponents obtained numerically and
analytically, which was also observed in the study of the
PEM~\cite{noh01}.

{\bf Acknowledgement:} We thank J.P. Bouchaud for a stimulating
discussion that motivated us to check the jump size statistics.  This
work has been supported financially by the Deutsche
Forschungsgemeinschaft (DFG).

\end{document}